\documentclass[article,twocolumn,showpacs,superscriptaddress,amsmath,amssymb]{revtex4}
\usepackage{cancel}
\usepackage{maybemath}
\usepackage{xcolor}
\usepackage{graphicx}

\begin{document}
published as Astropart. Phys., 99, 21-29 (2018)\\
\title{The Mont Blanc neutrinos from SN 1987A:\\Could they have been monochromatic (8 MeV)\\ tachyons with $m^2=-0.38$ keV$^2$?}

\author{Robert Ehrlich}
\affiliation{George Mason University, Fairfax, VA 22030}
\email{rehrlich@gmu.edu}
\date{\today}
\begin{abstract}
According to conventional wisdom the 5-hour early Mont Blanc burst probably was not associated with SN 1987A, but if it was genuine, some exotic physics explanation had to be responsible.  Here we consider one truly exotic explanation, namely faster-than-light neutrinos having $m_\nu^2=-0.38$ $keV^2.$  It is shown that the Mont Blanc burst is consistent with the distinctive signature of that explanation i.e., an 8 MeV antineutrino line from SN 1987A.  It is further shown that a model of core collapse supernovae involving dark matter particles of mass 8 MeV would in fact yield an 8 MeV antineutrino line.  Moreover, that dark matter model predicts 8 MeV $\nu,\bar{\nu}$ and $e^+e^-$ pairs from the galactic center, a place where one would expect large amounts of dark matter to collect.  The resulting $e^+$ would create $\gamma-$rays from the galactic center, and a fit to MeV $\gamma-$ray data yields the model's dark matter mass, as well as the  calculated source temperature and angular size.  These good fits give indirect experimental support for the existence of an 8 MeV antineutrino line from SN 1987A.   More direct support comes from the spectrum of $N\sim1000$ events recorded by the Kamiokande-II detector on the day of SN 1987A, which appear to show an 8 MeV line atop the detector background.  This $\bar{\nu}$ line, if genuine, has been well-hidden for 30 years because it occurs very close to the peak of the background.  This fact might ordinarily justify extreme skepticism.  In the present case, however, a more positive view is called for based on (a) the very high statistical significance of the result $(30 \sigma),$ (b) the use of a detector background independent of the SN 1987A data using a later K-II data set, and (c)  the observation of an excess above the background spectrum whose central energy and width both agree with that of an 8 MeV $\bar{\nu}$ line broadened by $25\%$ resolution.  Most importantly, the last observation is in accord with the {\emph {prior prediction}} of an 8 MeV $\bar{\nu}$ line based on the Mont Blanc data, and the the dark matter model, itself supported by experimental observations.  Lastly, it is noted that the tachyonic interpretation of the Mont Blanc burst fits the author's earlier unconventional $3+3$ model of the neutrino mass states.  Experimental corroboration should be sought for the linked hypotheses of an 8 MeV $\bar{\nu}$ line or an $m_\nu^2=-0.38$ $keV^2.$   The former might be seen in existing astrophysical data, while the latter should be proven or refuted by the KATRIN experiment in a short data-taking period.\\

keywords: neutrino mass, SN 1987A, supernova, dark matter, galactic center, tachyon, KATRIN
\end{abstract}
\maketitle

\section{Neutrinos from SN 1987A\label{sect2}} 
On February 23, 1987 bursts of a few dozen neutrinos and antineutrinos were seen in the four detectors then operating, the largest of which was in Kamiokande-II.~\citep{Hi1988}  Three of the bursts occurred about the same time, but the fourth 5-event burst seen in the small LSD (Mont Blanc) detector preceded the other three by about 282 min.~\citep{Aga1988, Agb1987}  Since the four detectors were unsynchronized, we make the usual assumption to let $t=0$ for the earliest arriving neutrinos seen in the 3 detectors other than LSD.  In addition to observing the arrival times of the neutrinos the detectors also measured their energies, $E_i,$ which could be deduced from the ``visible" (positron) energies, $E_{vis}$ based on $E_i=E_{vis}+1.3$ MeV, assuming the dominant reaction to be $\bar{\nu}_e+p\rightarrow n+e^+.$  

\subsection{The puzzling Mont Blanc burst}
The LSD burst has been puzzling for at least three reasons besides its early arrival:

1. \emph{the absence of the early burst in the other detectors.} 
The absence of the burst in the Baksan and IMB detectors is understandable in view of their high thresholds and/or small size, but its absence in the K-II detector might seem more problemmatic.   However, it is not clear that the LSD burst was in fact absent in the K-II detector, which apparently saw 4 pulses within the 10 s time interval encompassing the LSD burst –- a coincidence that should
occur at random only once in 17 years, according to LSD Collaborators.~\citep{Aga1988}
 
2. \emph{the absence of the main burst in the LSD detector.} The neutrinos in the main 10-15 sec burst have energies mostly above 8 MeV, making any difference in the K-II and LSD detector efficiencies relatively unimportant compared to the difference in their masses.  Thus, LSD with only $4\%$ the mass of K-II would not have seen the main burst if K-II detected only 12 events at that time.

3. \emph{the near equal energies of the 5 LSD events.}  The LSD neutrino energies are all consistent with $E_\nu=8.0$ MeV, given their estimated meaurement uncertainty of $\pm 15\%$ with a $\chi^2$ probability of $67\%.$  Others have come up with various exotic explanations of the LSD neutrinos,~\citep{DeR1987,Sc2015, Fro2016}.  Vissani, however, has noted that if they were not just a background fluctuation that no models exist to explain their virtually identical energies.~\citep{Vi2014}  As will be seen, both the constancy of their energies and the specific value of $E_{avg}=8$ MeV have a natural explanation given the hypothesis under consideration here.

\subsection{An $m^2<0$ neutrino and an $E_\nu=8$ MeV line}  We here ignore the theoretical difficulties posed by $v>c$ and $m^2<0$ neutrinos, which have been dealt with elsewhere,~\citep{Ma2016,Ch1985, Ci1999,Ra2010, Je2016, Je2017, Eh2012,Eh2013,Eh2015,Eh2016}, and assume they obey the usual kinematic equations of special relativity for an $m^2<0$ particle. The $\nu$ and $\bar{\nu}$ flavor states created in a supernova core collapse each consist of a mixture of mass eigenstates that lose their coherence en route to Earth from a distant supernova.  Essentially the wave packets of any pair of mass eigenstates no longer overlap and interfere after they have travelled a distance from SN 1987A greater than the coherence length.~\citep{Gi1998}  Moreover, for $|m|>>1 eV$ the spread in arrival  times of the SN1987A antineutrinos due to a spread in emission times will be dwarfed by a spread in travel times,~\citep{Eh2012} so one can identify the mass $m$ of $\emph {individual}$ $\bar{\nu}$ based on their arrival times $\Delta t_i$ and energy $E_i.$  Let us define $t_i$ and $\tau$ as the travel times for a neutrino and a photon, respectively, and assume $c=1.$  In the limit $m<<E_i$ we therefore find for the neutrino speed $v_i=\tau/t_i=1-m^2/2E_i^2,$ from which we immediately obtain this linear relation between $1/E_i^2$ and $\Delta t_i\equiv t_i-\tau$ 

\begin{equation}
\frac{1}{E_i^2}=\frac{2}{t_i m^2}\Delta t_i\approx \frac{2}{\tau m^2}\Delta t_i
\end{equation}
According to Eq. 1 the $1/E_i^2$ and $\Delta t_i$ coordinates for individual neutrinos having the same $m^2$ will lie on a line through the origin having a slope $2/\tau m^2.$  One can use Eq. 1 to solve for the mass eigenstate inferred from the LSD observation.  With $\Delta t = -282$ min and $E_{avg}=8.0$ MeV, one finds $m_{avg}^2 = -0.38$ keV$^2$ 
It is clear from Eq. 1 that one would expect a tachyonic neutrino with a $|m^2|$ as large as $0.38 keV^2$ will (almost always) not be seen as a single burst of neutrinos in a narrow time window, but instead they will lie on or close to a negatively sloped line, and likely be spread out in arrival time perhaps over many hours before $\Delta t = 0$ because of their spread in energy.  The only way that the burst could be observed as arriving in a narrow time window would be if the neutrinos all have nearly the same energy.  Given the observation time for the burst $|\Delta t| = 282$ min$ = 16,900$ sec and its duration $\delta(\Delta t) =7s$, if we assume the 5 events are all associated with a common mass $m^2,$ one can infer a maximum spread in the energies of the five events based on Eq. 1 as

\begin{equation}
\frac{\delta E}{E}=\frac{\delta (\Delta t)}{2|\Delta t|} =  \frac{7 s}{2\times 16,900 s} =\frac{1}{4834}\approx 0.02\%
\end{equation}
which is essentially a line in the (anti)neutrino spectrum.  In order for the $m^2 \approx -0.38$ keV$^2$ explanation of the LSD burst to be remotely plausible there needs to be some core collapse model of SN 1987A that includes an antineutrino flux component that is monochromatic having $E_\nu\approx 8$ MeV.  There exists no known SN model for such a $\bar{\nu}$ line, so one is now proposed.  Lest the reader believe that this model has been simply proposed on an ad hoc basis to fit the desired result, we first spell out the empirical justification for the model, and then consider the empirical evidence supporting it.

\section{SN Model with an $E_\nu= 8$ MeV line\label{sect3}}
Many researchers have previously suggested that dark matter $X$ particles might collect in the core of some stars by gravitational attraction and subsequently annihilate.~\citep{Ba2008}  Calculations based on full nonlinear simulations of the field equations show that the accumulated DM particles in the core could increase until some peak value is reached close to the Chandrasekhar limit.~\citep{Br2015}.  

Without some ``extra" energy source such as that provided by dark matter annihilation the stalling of the shock wave has in the past proven to be a severe difficulty with most models. Although current 3-D models incorporating non-spherically symmetric explosions are said to be``within reach" of solving the stalling problem, as of late 2017, one of the best of these models (s18-3D) is acknowledged to have elements that ``can still only be understood in qualitative terms," and so it represents at best ``only a step towards a solution of the problem of shock revival ..."~\citep{Ja2017b} 

\subsection{How to have an 8 MeV antineutrino line}
For a pair of dark matter $X$ particles to be converted to $\nu\bar{\nu}$ during their annihilation suggests the existence of some mediator or force carrier particle ($Z'$), as in the reaction: $XX\rightarrow Z'\rightarrow \nu\bar{\nu}$ which we henceforth refer to as the $Z'_\nu$ reaction.  Similarly, should the final product be $e^+e^-$ instead of $\nu\bar{\nu}$ we have the $Z'_e$ reaction.  Essentially, the hypothetical $Z'$ would be a mediator particle that serves as a portal linking DM $X$ particles to Standard Model leptons.  Clearly the existence of such a $Z'$ cannot be accommodated within the standard model, and such a possible ``fifth force" has long been of interest both experimentally and theoretically.~\citep{Fr2016}

In 2016 an anomaly was found for internal $e^+e^-$ pair creation in the reaction $^7Li(p,\gamma)Be^8.$~\citep{Kr2016}  By examining the number of pairs with various opening angles, $\theta,$ Krasznahorkay and collaborators found a $6.8\sigma$ excess for $\theta=140^0,$ yielding a chance probability of $p=5.6\times 10^{-12}.$  
This bump was interpreted as evidence for an intermediate short-lived particle with mass $m=16.7\pm 0.6$ MeV appearing in the two step decay process of the excited $^8Be,$ i.e.: $^8Be^*\rightarrow ^8Be Z',$ followed by $Z'\rightarrow e^+e^-$ with a branching ratio about $5\times 10^{-6}$ that of $^8Be^*\rightarrow ^8Be\gamma$.  Feng et al.~\citep{Fe2016} have presented evidence that this new particle, a gauge isoscalar boson, is the mediator of a fifth force having a short (12 fm) range, and Chen et al.~\citep{Ch2016} have shown that it couples almost exclusively to $e^+e^-$ and $\nu\bar{\nu},$ and that they suggest that it is the force carrier linking dark matter $X$ particles to those of the Standard model through the $Z'_\nu/Z'_e$ reactions.  

It is important to note that the $\nu\bar{\nu}$ created in $XX$ annihilations would be nearly monochromatic with energy $m_X$ as long as the $X$ particles are relatively cold, and hence cause little Doppler broadening.  Finally, given cold X particles ($v_X/c<<1$), and $Z'_\nu/Z'_e$ reactions taking place on resonance (for the largest cross section) we must have $m_X\approx \frac{1}{2}m_{Z'},$ making the end product of the reaction nearly monochromatic $\nu$ and $\bar{\nu}$ pairs having $E_\nu=8.4\pm 0.3$ MeV.  The recently discovered 16.7 MeV $Z'$ particle is therefore the ideal candidate for creating 8 MeV neutrino and antineutrino lines from $XX$ annihilation via the $Z'_\nu$ reaction.  Further experiments are essential to verify the Krasznahorkay et al. result, but as of early 2018 there are no experiments in conflict with it.

\subsection{The $Z'_e/Z'_\nu$ mediated reaction model\label{flux}}
It is noteworthy that a model involving dark matter for producing monochromatic $\nu,\bar{\nu}$ pairs via the see-saw mechanism already exists,~\citep{Du2015} however, that model was not proposed in connection with supernovae, nor did it suggest any particular neutrino energy.  The model being proposed here for  8 MeV $\nu$ and $\bar{\nu}$ lines from Type II SNe incorporates the Chen et al. proposal that the 16.7 MeV $Z'$ particle can act as a portal (in both directions) between DM $X$ particles and Standard Model (SM) leptons.  The initial stage of a core collapse is fueled by the release of $\nu_e$ mostly from electron capture reactions, just as in conventional models, and the infalling SM matter heats the core very rapidly.  Due to the initial absence of interaction between DM $X$ particles and SM particles they can be at two very different temperatures, with the DM initially assumed to follow a Maxwell-Boltzmann distribution.~\citep{Lo2011}  However, once the core collapse proceeds to the point where the SM particles achieve some (very high) threshold temperature the SM-DM portal is partly breached from the SM side, and the inverse $Z'_e$ reaction ($Z'_e(inv)$) can occur:  $e^+e^-\rightarrow Z'\rightarrow XX.$  This creates many``hot spots" in the DM above the threshold for the $Z'_\nu$ reaction resulting in energy flows in both directions, with a rapidly increasing number of reacting particles.  Shortly, the SM-DM portal is fully open, and the dominant flow of energy is from the low threshold DM side -- which we may think of as the ``high water side of the breached  dam." Those $\nu\bar{\nu}$ pairs produced only slightly above the $Z'_\nu$ threshold will constitute a spectral line emitted during the initial short burst, but there will also be a non-monochromatic component having  energies significantly above threshold.  In fact it could be some time before the DM hot spots cool (via the $Z'_e/Z'_\nu$ mediated reactions) to being just above threshold, and this raises the possibility of a monochromatic component of $\nu\bar{\nu}$ that is emitted both in a brief burst, and over a time scale much longer than 10 s -- important in connection with the claim made in the appendix.
\begin{figure}[t!]
\centerline{\includegraphics[angle=0,width=1.0\linewidth]{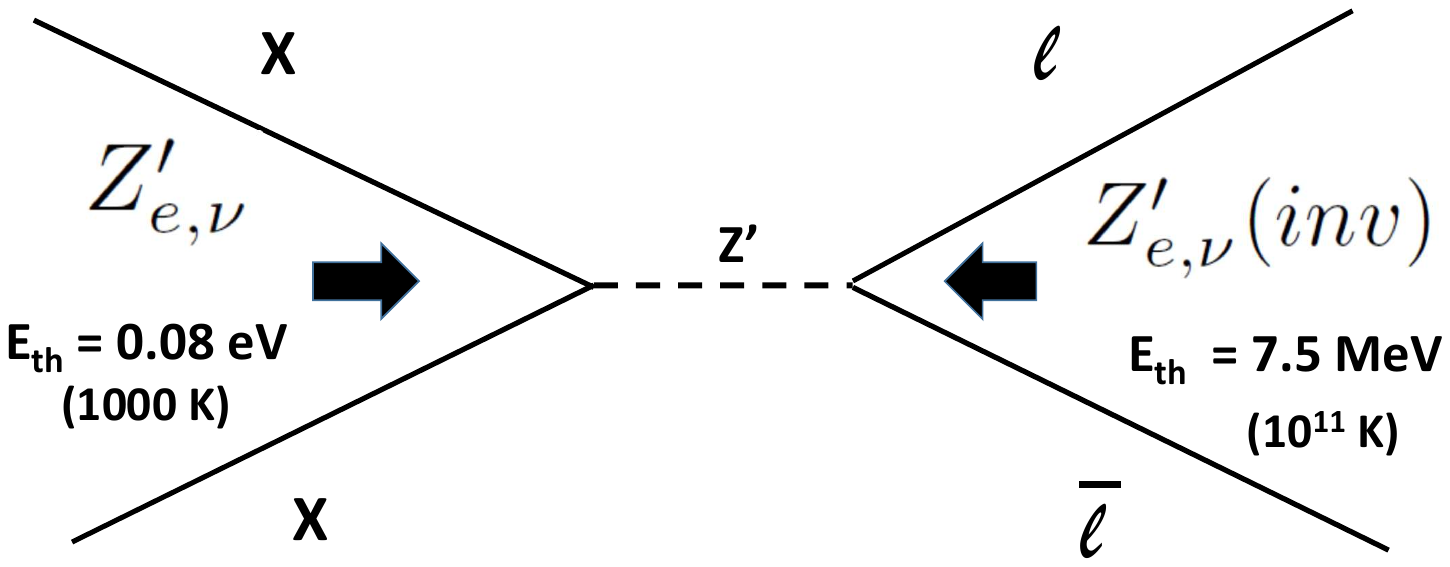}}
\vspace{-0.75in}
\caption{\label{fig3}{Diagram of the direct and inverse $Z'_e/Z'_\nu$ mediated reactions for $XX$ annihilation into leptons, and the reaction thresholds in each direction suggested in the text.  7.5 MeV is the threshold for $e^+e^-,$ while that for $\nu\bar{\nu}$ is 8 MeV.}}
\end{figure}
\subsection{The model values of $T$ and $m_X$}
The threshold kinetic energy of each $X$ in the CM frame of the core can be expressed as: $E_{th}\equiv T=\frac{1}{2}m_{Z'}-m_X.$  A further constraint can be set by the requirement that in order that the emitted $\nu \bar{\nu}$ be highly monochromatic we must have the velocities $v_X$ and $v_{Z'}$ (relative to the core CM frame) obey $v<<c$ or $E_{th}<<m_X,$ from which it follows, as noted earlier, that the X-particle mass is $m_X\approx \frac{1}{2}m_{Z'} = 8.4\pm 0.3$ MeV.  We can now estimate a value for $E_{th}$ based on the degree of monochromaticity of the neutrinos in the Mont Blanc burst based on Eq. 2, which assumed they really were due to a single tachyonic mass state, i.e. $\Delta E/E\approx 0.0002.$  Now consider emitted $\nu \bar{\nu}$ pairs in the $Z'$ rest frame each having $E=8$ MeV.  Based on the standard transformation equations we find that in the core CM frame we have to first order in $v_{Z'}$ that  $E_\pm=(1\pm v_{Z'})E,$ where $\pm$ signs indicate forward or backward emitted neutrinos in the $Z'$ rest frame.  The spread in energy between these two extremes satisfies:  $\Delta E = 2v_{Z'}E\approx 0.0002E,$ so that $v_{Z'}\approx 10^{-4}c.$  The rms average of the $X$ and $Z'$ velocities are related through $v_{Z'rms}=v_{Xrms}/\sqrt{2},$ from which we conclude that the threshold kinetic energy for the $Z'_\nu$ and $Z'_e$ reaction is $E_{th}=\frac{1}{2}m_Xv^2_{Xrms}\approx  0.08$ eV or $T\approx 1000K.$

\section{Testing the DM model\label{sect4}}
The $Z'_e/Z'_\nu$ mediated reactions would occur not just in the case of core collapse SNe, but anywhere there is a large concentration of $X$ particles and a sufficiently high temperature to reach the $Z'_e/Z'_\nu$ threshold, and the galactic center is an obvious possibility.  It will be shown in this section that the DM model gives an excellent fit to observations of MeV $\gamma-$rays from the galactic center, which supports the reality of an 8 MeV antineutrino line from SN 1987A.

\subsection{Gamma rays from the galactic center}
The $Z'_e$ reaction will result in monochromatic $e^+e^-$ pairs, and when the $e^+$ annihilates at rest the result is a 511 keV gamma-ray.  According to Ref.~\citep{Wi2016}, the high luminosity of such emissions from near the galactic center is not explained by any known mechanism, and the cuspy spatial shape is highly suggestive of DM annihilation.   Those same authors, however, argue against this possibility, because they assert that a DM explanation of the 511 keV emission is strongly disfavored on theoretical grounds in the simplest of models, i.e. thermal production with no extra particles.~\citep{Wi2016}  Of course the proposed $Z'_e/Z'_\nu$ mediated reaction model, does involve an ``extra" particle ($Z'$) for which there is good empirical evidence, so the Wilkinson et al. theoretical objections to a DM origin of 511 keV emissions do not apply.   Ajello et. al has also suggested that the DM explanation of GC $\gamma-$rays is no longer tenable.~\citep{Aj2017}  While those observations based on Fermi-LAT multi-GeV data might rule out a proposed DM particle having $m_X$ in that energy range, they have nothing to do with a possible DM origin of MeV $\gamma-$rays.

\begin{figure}[t!]
\centerline{\includegraphics[angle=0,width=1.1\linewidth]{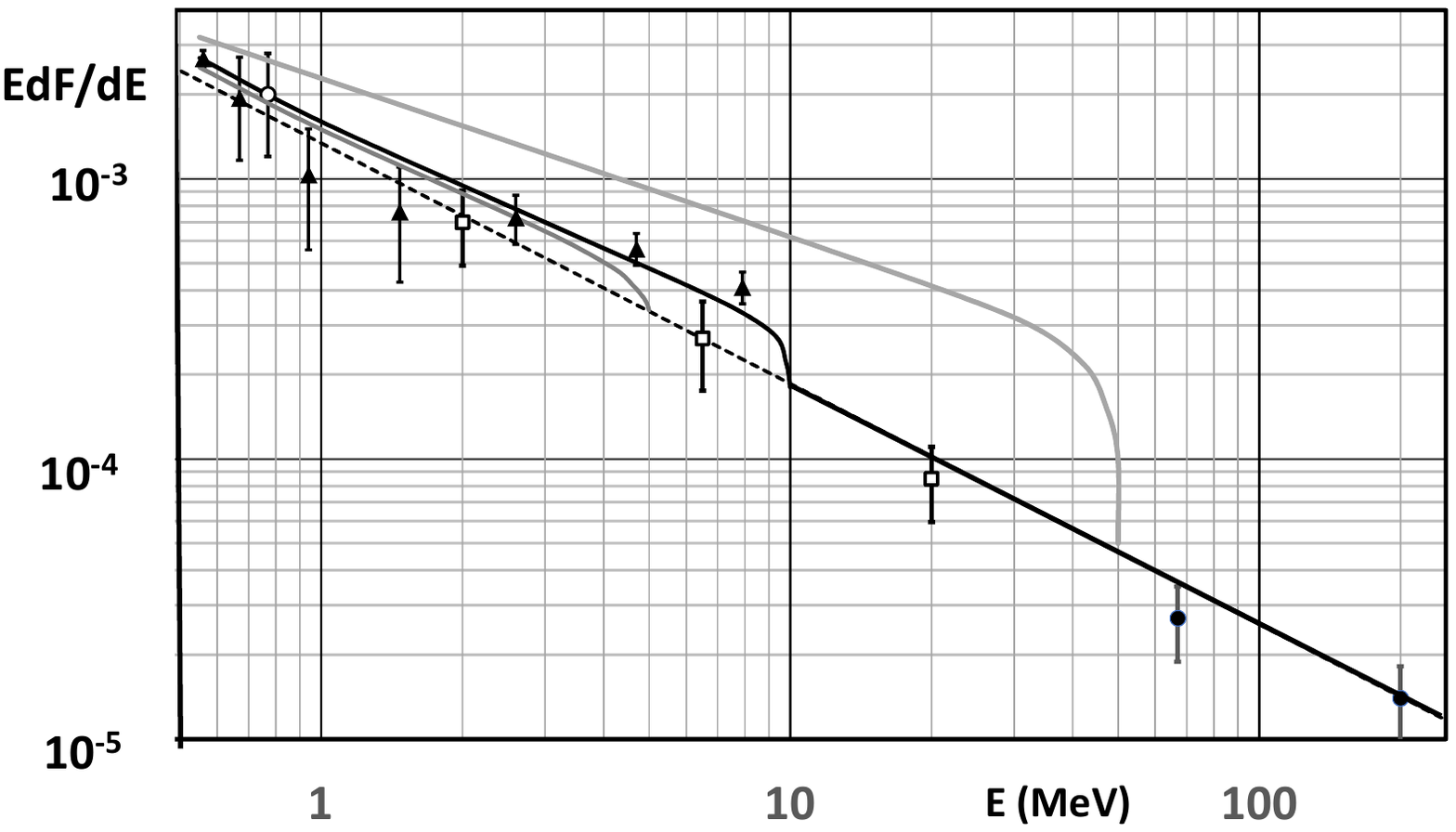}}
\vspace{-0.35in}
\caption{Spectrum, i.e., $E\times\frac{dF}{dE} (cm^{-2}s^{-1})$ versus energy for $\gamma-$rays from the inner galaxy for $E>511$ keV, as measured by 4 instruments: SPI(open circle), COMPTEL (open squares), EGRET (filled circles), and OSSE (filled triangles).  All but the 7 OSSE points (from Ref.\citep{Ki2001}) are from Prantzos et al.~\citep{Pr2010}, as are the 3 predicted enhancement curves above the straight line for positrons injected into a neutral medium at initial energies $E_0=5, 10, 50$ MeV displayed as the lower grey curve, the black curve, and the upper grey curve, respectively.  The sloped straight line (also from Ref.~\citep{Pr2010} is a power law fit to the spectrum at high and low energies.}
\end{figure}

\subsubsection{The positron source temperature}
The temperature $T$ of the medium from which the $e^+$ originated affects the degree of Doppler broadening of the 511 keV line, $\Delta E.$  However, two other poorly known variables also affect $\Delta E,$ i.e., the positronium fraction, $f_{pos}$ formed before annihilation, and the fraction of the medium that is ionized, $f_{ion}.$  Doing a four parameter simulation, Churazov et al.\citep{Ch2005} have found that given the measured values of $\Delta E$ and $f_{pos}$ two solutions are possible, a ``warm" one with $7000K < T < 40,000K,$ and a ``cold" one with $T\leq 1000K.$   Note that the cold solution is consistent with the previous estimate of the $T\approx 1000 K$ threshold for the $Z'_e$ and $Z'_\nu$ reactions, and additionally the SPI collaboration notes that their measurements are also consistent with $T\approx 1000K,$ within uncertainties.~\citep{Si2015}
  
\subsubsection{Finding $m_X$ by fitting the $\gamma-$ray spectrum}
Most monochromatic $e^+$ produced in the $Z'_e$ reaction will form positronium and essentially annihilate at rest.  It would be only the small fraction  $1-f_{pos}=3\pm 2\%$~\citep{Je2006} of these $e^+$ that annihilate in flight that could give information on their original total energy, $E_0= m_X.$  Those $3\%$ of the $e^+$ that annihilate in flight cause a broad enhancement to the $\gamma-$ ray spectrum above the 511 keV line whose shape can be calculated based on the rate of $e^+$ energy loss in the interstellar medium, and the fraction of the medium that is assumed to be ionized, $f_{ion}$.  Fig. 2 shows the predicted enhancement for three values of $m_X$ in the case of $e^+$ propagating in a neutral medium ($f_{ion}=0\%$), together with data from four instruments.  It may be noted that the enhancement in each case drops to zero, i.e., the spectrum joins the straight line power law at an energy $E=m_X.$  This result follows from conservation of energy for those $e^+$ that annihilate in flight immediately with no energy loss.  It is clear that the data are inconsistent with the depicted enhancement curves for $m_X =50$ MeV and $m_X=5$ MeV (both grey), but they are consistent with $m_X=10$ MeV (black).  In fact $m_X=10$ MeV, which gives a very good fit to the data $\chi^2=7.3,$ $(p=89\%, dof=13),$ is very close to being a best fit.  Moreover acceptable fits can only be found for the range: $m_X=10^{+5}_{-2}$ MeV, which is in excellent agreement with the $m_X$ mass in the $Z'_e/Z'_\nu$ mediated reaction model, i.e., $m_X=8.4\pm 0.3$ MeV.  It is important to note that the data and fitted curves in Fig. 2 first appeared in Prantzos et al.,~\citep{Pr2010} with the exception of the 7 data points from the OSSE instrument taken from Ref.~\citep{Ki2001} that were added by the author.  In the absence of the OSSE data Prantzos et al. concluded that their graph constrained $m_X$ to being``less than a few MeV."~\citep{Pr2010}  Regardless of the questionable merits of that claim, it is clear that the inclusion of the OSSE data changes the situation dramatically, in view of the small error bars on four of those points.  

\subsubsection{Why the small OSSE error bars?}
The upper and lower limits on the  value of $m_X$ are due to the very small error bars of three of the OSSE data points (one at $E=0.56MeV$ and the two for $E>4 MeV$ that lie about $4.5\sigma$ and $6.0\sigma$ respectively above the dotted line).  One might be puzzled as to why some of the OSSE data points have such small error bars compared to those from the other instruments.  For the other three instruments whose data is depicted the source acceptance consisted of a broad area centered on the galactic center having a size $(\Delta l=\pm 10^0, \Delta b=\pm 10^0),$ whereas for OSSE a much smaller acceptance area up to $(\Delta l=\pm 3^0, \Delta b=\pm 3^0)$ was used.  In both cases the excess flux from the galactic center is computed after subtracting the off-source flux, and if the area occupied by the source is in fact very small (as it is now known to be: $\sigma(\theta)\approx 2.5^0$), one gets a minimum error by using an acceptance for the on-source measurement that is only slightly larger than the source itself.  

\subsubsection{Observed and predicted $\gamma-$ray source radius}  
A supermassive black hole known as $Sgr A^*$ having $M=4\times 10^6m_\odot$ is located near the GC $(l=-0.06^0,b=-0.05^0),$ and gas falling into $Sgr A^*$ attains temperatures as high as $10^{10} K$ or more,~\citep{Me2001} the threshold for pair creation.  The radial distribution of temperature in the accretion disk outside a supermassive black hole of mass $M$ and mass accretion rate $\dot{M}$ is given by~\citep{Fr2002}:
\begin{equation}
T(R)=\left[\frac{3GM\dot{M}}{8\pi\sigma R_S^3}\right]^{1/4}\left(\frac{R}{R_S}\right)^{-3/4}
\end{equation}
where $R_S=2GM/c^2$ is the Schwarzschild radius, and it is assumed $R>>R_S.$  Consider two spheres of radii $R_1$ and $R_2$ concentrically surrounding $Sgr A^*,$ whose surfaces are respectively at temperatures $T_1$ and $T_2.$ By the previous equation we have: $R_2=R_1\left(T_1/T_2\right)^{4/3}.$  Thus, suppose the inner sphere of radius $R_1=0.1''$, where the temperature is known to be $T_1=5\times 10^6K.$~\citep{Me2001}.  We find that if the outer radius sphere has a temperature equal to the model-predicted threshold $T_2=1000 K,$ then $R_2=0.1''\times\big({\frac{5\times 10^6}{1000}\big)^{4/3}}=2.4^0,$  in excellent agreement with the measured angular radius of the GC $\gamma-$ray source, i.e., $2.5^0.$~\citep{Pr2010} 

\begin{table}[h]
\centering
\begin{tabular}{lcc}
\hline\hline
Quantity &\hspace{0.25in} observed value &\hspace{0.25in} predicted value \\ 
\hline
$m_X$ & $10^{+5}_{-2}MeV$ & $ 8.4\pm 0.3 MeV$\\
$\sigma(\theta)$  & $2.5^0$   &  $2.4^0$\\
$T$                   & $10^3K$ & $10^3K$\\
\hline
\end{tabular}
\caption{Values of $m_X,$ T, and $\sigma (\theta)$ from observations and predictions by the $Z'_e/Z'_\nu$ mediated reaction model.}
\end{table}

\section{Possible Validation}
The two ways to validate the exotic hypothesis of this paper would be to seek conclusive evidence for either an 8 MeV $\bar{\nu}$ line or alternatively a neutrino having $m^2=-0.38$ keV$^2.$

\subsection{Search for an 8 MeV $\bar{\nu}$ line}

Once the next SN occurs in our galaxy, the presence or absence of an 8 MeV line should be quite obvious given the sensitivity of today's neutrino detectors.~\citep{Hi2016}   Some detectors such as Borexino or LVD~\citep{Pa2017,Br2017} do have low enough thresholds to detect an 8 MeV line from not yet observed relic (diffuse) SNe or from the galactic center, but the issue of high background at this energy is very serious.    According to Beacom and Vagins,~\citep{Be2004}, the flux from diffuse SNe at $E_\nu=8$ MeV might be as much as two orders of magnitude below that of the background $\bar{\nu_e}$ from nuclear reactors -- a value they estimated based on Super-Kamiokande data.   This very high background might seem to exclude the possibility of looking for an 8 MeV $\bar{\nu}$ line in such data.  However, we should recall that given the inverse square law, the nuclear reactor background does depend on the reactor distance, and Super-K happens to be very poorly located in this respect, with the nearest reactor only about 100 km away.  In contrast, were one to consider a neutrino detector located 4000 km from the nearest reactor, we estimate an approximate background reduction from reactor $\bar{\nu_e}$ by over three orders of magnitude compared to that for Super-K, making any diffuse SN neutrinos much more observable than suggested by Beacom and Vagins.  Moreover even when there are reactors near a detector, one can reduce the reactor background considerably by excluding $\bar{\nu_e}$ having arrival directions $\theta$ from any reactor closer than 1000 km, within a cone of half angle $\Delta\theta_{res}.$    Virtually the only other background at 8 MeV, due to solar $\bar{\nu_e},$ can also be greatly reduced by making a similar exclusion on arrival directions near the sun.  Still further reductions in background are achievable using a novel search method proposed by Casentini et al.~\citep{Ca2018}.  The preceding considerations suggest that the search for $\bar{\nu}$ from diffuse SN or the galactic center need not be limited to $E_\nu>11 MeV,$ as sometimes assumed.~\citep{Lu2016}  Thus, it would be most interesting to examine existing data for low threshold detectors after making the appropriate  $\bar{\nu}$ arrival direction cuts to see if there were evidence of an 8 MeV $\bar{\nu_e}$ line above background, where for the background spectrum one might use that of the excluded events.  Finally, if such a peak were found, one should examine how its prominence depended on arrival directions with respect to the galactic center.  Of course, the absence of an 8 MeV $\bar{\nu}$ line in these data would not be conclusive evidence against the hypothesized $m^2=-0.38$ keV$^2$ neutrino, but the second proposed test below should offer a definitive proof for or against it.

\subsection{A $\bf{m^2_\nu=-0.38}$ $\bf{keV^2}$ neutrino and the $3+3$ model}
\begin{figure}[t!]
\centerline{\includegraphics[angle=0,width=1.0\linewidth]{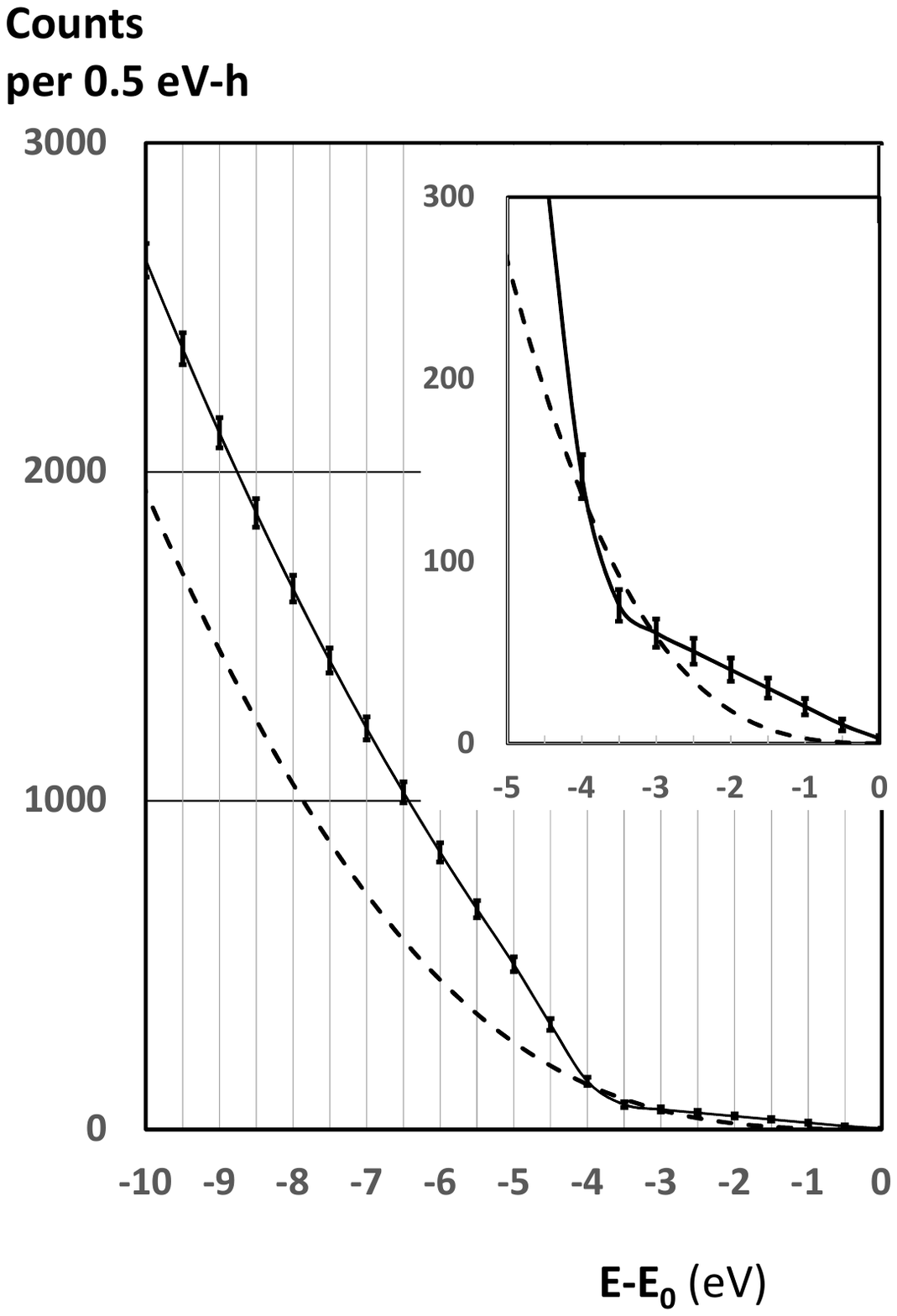}}
\caption{\label{fig3}Simulated KATRIN data based on the $3+3$ model masses for the last 10 eV of the spectrum.  The simulation is based on one hour of data-taking for each of 24 energy bins of 0.5 eV width normalized to yield the expected count rate at $E_0-E=20$ eV.  The dashed curve shows the all $m_j=0$ case after final state distributions and energy resolution have been included.  The insert shows the last 5 eV with an expanded vertical scale.}
\end{figure}
The tachyonic mass value $m^2_\nu=-0.38$ $keV^2$ is within a factor of two of the hypothesized value $m^2 \approx -0.2$ keV$^2$ 
originally postulated in a 2013 paper on the $3+3$ model, which made no use of the Mont Blanc data.~\citep{Eh2013}   This original value was only approximate, however, because it was based on the estimated $\Delta m^2_{sbl}\approx 1$ eV$^2$ for the large $\Delta m^2$ oscillation claimed in short baseline experiments, which in fact is uncertain by over a factor of two.~\citep{Gi2013}  

Evidence for the three masses in the $3+3$ model can be sought in direct mass experiments based on the shape of the $\beta-$decay spectrum.  In the region near the its endpoint $E=E_0,$ the spectrum can be well-approximated by the square of the Kurie function (the phase space term) for  multiple non-degenerate neutrino masses $m_j$:
\begin{equation}
K^2(E)=(E_0-E)\sum |U_{ej}|^2\sqrt{(E_0-E)^2-m_j^2}
\end{equation}

where the quantity inside the square root is replaced by zero if it is negative (which never happens when $m_j^2<0.$)   According to Eq. 4 the three spectral features due to the $3+3$ model masses: $m_1=4.0$ eV, $m_2=21.4$ eV and $m_3^2=-0.2$ eV$^2$ are:  a kink at (a) $E_0-E=21.4$ eV, (b) a second kink at $E_0-E=4.0$ eV, and (c) a linear decline in the last 4.0 eV.  Owing to limited statistics and resolution, existing experiments have revealed only the most prominent feature predicted by the model, i.e., the first kink.~\citep{Eh2016}   Some of those pre-KATRIN experiments, however, have chosen to plot the residuals to a fit to the spectrum using $m_\nu=0$ rather than the spectrum itself.  In this case, the kink shows up instead as an artifactual spectral line near the endpoint or a best fit $m_\nu^2<0$ mass value.   Although an alternate explanation for the observed kink at $E_0-E\approx 20$ eV involving possible electronic excitations in the tritium gas molecules~\citep{Bo2015}, cannot be excluded, we note that the $3+3$ model gives a significantly better fit to the data than the all $m_j=0$ case despite the limit $m_\nu\rm{(eff)}<2$ eV on the $\nu_e$ effective mass.  Fortunately, the KATRIN experiment~\citep{Dr2013} has the statistical and systematic sensitivity to make the definitive test of the model by observing all three features in the spectrum associated with the three $3+3$ model masses.  Thus, in particular Fig. 3 shows what KATRIN should observe in the last 10 eV before the spectrum endpoint when final state distributions, energy loss and resolution are included.  

\section{Summary}
In summary, the tachyonic interpretation of the Mont Blanc burst is shown to require the existence of an 8 MeV $\bar{\nu}$ line from SN 1987A.  A new $Z'_{e}/Z'_\nu$ mediated reaction model of the core collapse based on annihilating $m_X=8 MeV$ dark matter (a mass value suggested by a newly discovered 16.7 MeV boson) is found to predict such a line.  This model also predicts 8 MeV $e^+e^-$ pairs from the galactic center, and as Table I shows it agrees with three observations of $\gamma$-rays from that source, thereby providing both theoretical and empirical support to the reality of the $\bar{\nu}$ line.   In an appendix (because of its length), additional important evidence for an 8 MeV $\bar{\nu}$ line is presented based on an analysis of the Kamiokande-II data $(N\sim 1000$ events) taken on the day of SN 1987A in the hours before and after the main burst.
The evidence for the $m^2=-0.38$ keV$^2$ neutrino in the present paper is consistent with the previously published $3+3$ model,~\citep{Eh2013} which included a $m^2\approx-0.2$ keV$^2$ neutrino, and it adds to that earlier evidence: (a) good fits found in 3 high precision tritium $\beta-$decay experiments,~\citep{Eh2016}, (b) good fits to dark matter profiles for the Milky Way and and four galaxy clusters,~\citep{Ch2014} and (c) good fits to a $m^2<0$ value for the $\nu_e$ {\emph{flavor}} state from various sources of data.\citep{Eh2015}  Thus, until KATRIN yields its result, which should unambiguously prove or refute the existence of the tachyonic mass (and the two others) in the $3+3$ model, it could be a mistake to dismiss tachyonic neutrinos as being ``unphysical."~\citep{Lo2001,Kr2005} While it is certainly true that extraordinary claims require extraordinary evidence, it is also true that ``the eye sees only what the mind is prepared to comprehend."~\citep{Da2007}

\section{Appendix}
{\bf{Evidence for an 8 MeV $\bar{\nu}$ line from SN 1987A.}}\\

The idea of an 8 MeV antineutrino line from SN 1987A is contrary to all exisiting SNe core collapse models,~\citep{Ja2017a, Ja2017b} although the new dark matter model presented in this paper requires it, as previously discussed.  Here it is shown that two sets of pubished data taken by the Kamiokande-II Collaboration before and after SN 1987A can be interpreted as providing evidence for just such an $\bar{\nu}$ line. 

In their 1988 paper on SN 1987A  the K-II Collaboration provided data taken during eight 17 min-long time intervals in the hours before and after the 12-event burst.~\citep{Hi1988}   Figs. 4 (a)-(h) of Ref.~\citep{Hi1988} shows scatter plots  for each recorded event displaying the number of ``hits," $N_{hit},$ (PMT's activated) versus the event occurrence time during that $\Delta t = 8\times 17\rm{min}=0.094$ day time interval.  As can be seen in Fig. 4(a) the relation between $N_{hit}$ and $E_e,$ is quite linear, $E_e=cN_{hit}$ with $c=0.363\pm 5\%$ for the 12 events in the burst.

\begin{figure}[t!]
\centerline{\includegraphics[angle=0,width=1.0\linewidth]{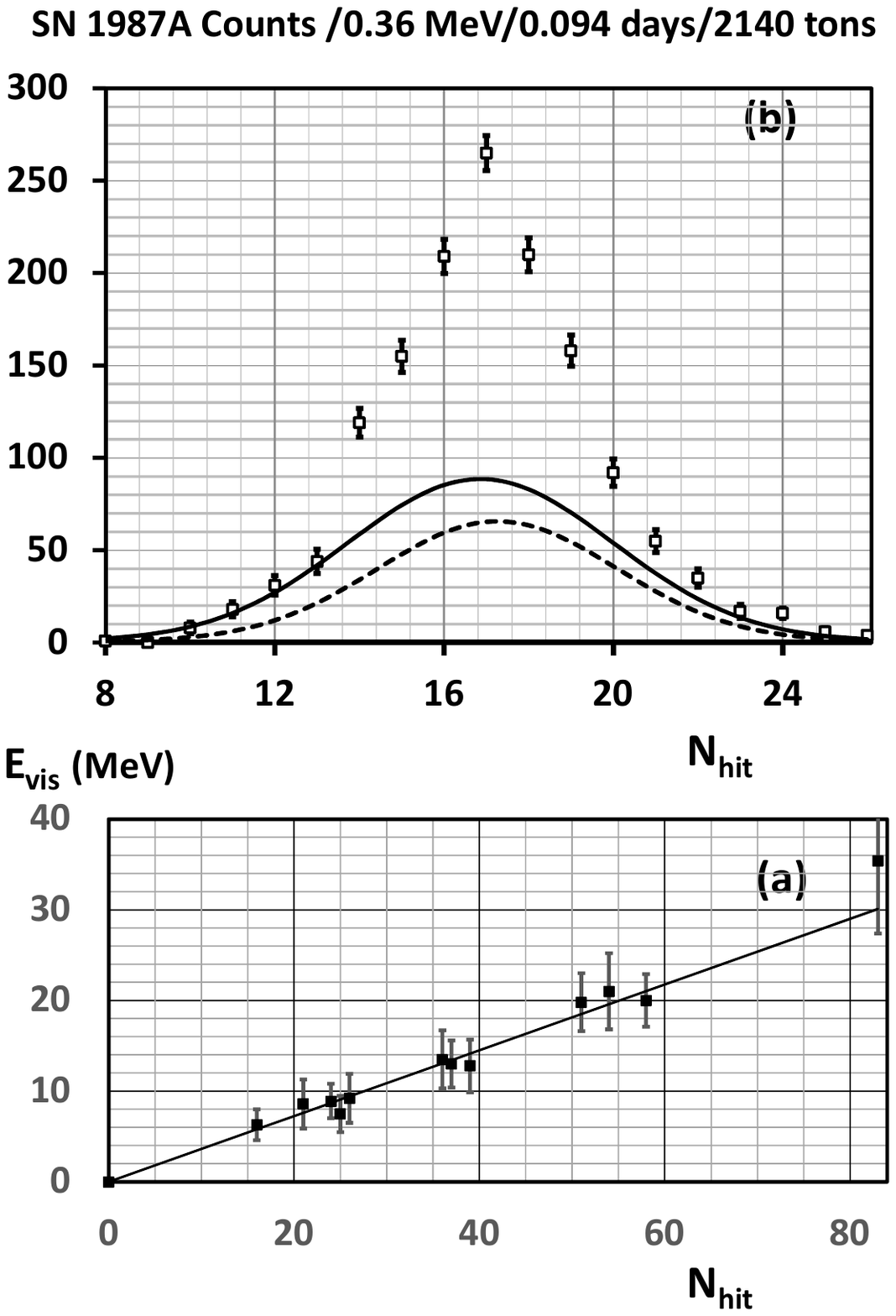}}
\caption{\label{fig3} (a) $E_e$ (total visible energy) versus $N_{hit}$ for the 12 events in the burst seen by K-II data for SN 1987A. (b) Histogram of $N_{hit}$ values for all events in Figs. 4 (a)-(h) in Ref.~\citep{Hi1988} The solid curve is the background for the detector that was found from a search for $^8B$ solar neutrinos -- see text.  The dotted curve shows an ``adjusted" background (see sect.\ref{triggers}), which is reduced in height by $26\%,$ decreased in width by $10\%,$ and shifted by $\Delta N_{hit}=+0.4.$}
\end{figure}
\begin{figure}[t!]
\centerline{\includegraphics[angle=0,width=1.0\linewidth]{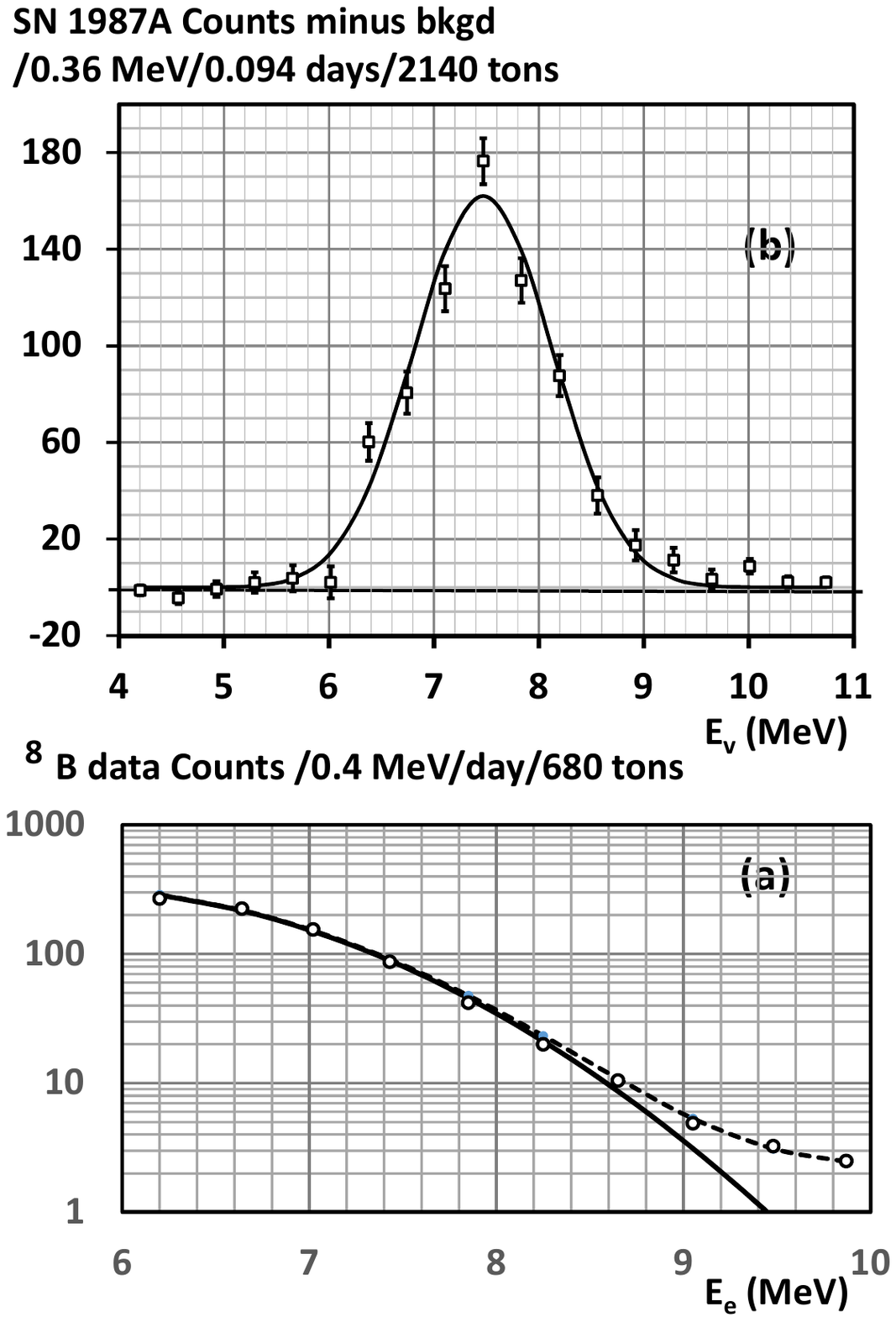}}
\caption{\label{fig4} (a) Data (circles) from a search for $^8B$ solar neutrinos.  The dashed curve is our fit to the $^8B$ data after a 680T fiducial volume cut.  (b) The background subtracted histogram for the SN 1987A data and a Gaussian fit to it after changing the horizontal scale to $E{_\nu}=0.363N_{hit} +1.3 MeV.$  The unadjusted solid background has been used for this plot.}
\end{figure}

The histogram in Fig. 4(b) was generated based on counting the number of times various $N_{hit}$ values were found in Figs. 4 (a)-(h) in Ref.~\citep{Hi1988}.  The question of whether the depicted $\sim 600$ event excess above background has physical significance should be viewed with obvious skepticism, since the maximum $N_{hit}$ count falls right at the peak of the claimed background distribution, and the excess above background is not confined to a narrow region of $N_{hit}$ or $E_\nu.$  Attributing physical significance to the excess clearly requires that the background has been obtained independent of the SN 1987A data, and that both its shape and normalization are known to an uncertainty that is small compared to the size of the excess.  It must also be shown that the distribution of the excess counts closely matches what is expected for a spectral line.
\subsection{Finding the SN 1987A background}
In 1989 the K-II Collaboration published the results of a search for solar neutrinos from the reaction ${^8}B\rightarrow {^8Be^*}+e^++\nu_e$ based on 450 days of data.~\citep{Hi1989}  The beginning of the data-taking period overlapped the date of SN 1987A, but most of it was many months afterwards.  Generally, models of neutrino luminosity show rapid power law declines with time during the SN cooling period.  As a result, the number of neutrinos observed months after the main burst that were due to the SN will be a far smaller fraction of the background than for a time interval that was only minutes or hours after the main burst.  This fact, {\emph{in principle,}} allows the $^8B$ data to define the SN background spectrum, and to do so with very little statistical uncertainty, given the factor of 5000 difference in the observation times for the $^8B$  and SN data sets.  The $\nu_e$ from $^8B$ are detected via the reaction $\nu_e+e^-\rightarrow\nu_e+e^-,$ unlike the $\bar{\nu}_e$ from SN 1987A are detected mainly by the $e^+$ they create.  However, the background spectra in both data sets due to radioactivity and cosmic rays are effectively the same because the sign of the charge of the $e^\pm$ is not distinguished by the measured quantities.~\citep{Hi1989}

Fig. 5(a) displays data extracted from Fig. 1 in Ref.~\citep{Hi1989}, and it shows (as circles) the distribution of $e^+/e^-$  energies in the $^8B$ data set for $E_e>6 MeV$ after a 680 ton fiducial cut was made.   Also shown in Fig. 5(a) as a dashed curve is our fit to the $^8B$ data consisting of a Gaussian plus a constant: $N = Ae^{(E-E_0)^2/2\sigma^2}+C,$ which is what is used to obtain the background curve for the SN data.  However, there must be an adjustment to the values of $A, E_0, \sigma$ and C between the curves displayed in Figs. 5(a) and 4(b) based on the different units indicated by the top labels in those figures.   Thus, for now the background for the SN data is assumed to be the same as the $^8B$ data after adjustments due {\emph {only}} to the different time intervals, fiducial masses, and energy bin widths in the two data sets.  Later a further adjustment to the SN background is shown to be needed -- see dotted curve in Fig. 4(b).  This background adjustment due to slightly different triggering criteria for the two data sets can be seen to lead to a relatively small change in the size of the $\sim 600$ excess counts integrated over energy.

Fig. 5(b) shows the result of a background subtraction for the histogrammed SN data in Fig. 4(b), after converting the horizontal scale from $N_{hit}$ to the {\emph {neutrino}} energy $E_\nu=cN_{hit}+1.3$ MeV.  Given the $5\%$ uncertainty in $c=0.363$, the fitted Gaussian in Fig. 5(b) is found to center at $E_{max}=7.5\pm 0.4$ MeV, and have $\Delta E/E_{max}=21\%.$  The latter value is consistent with the expected $25\%$ energy resolution based on $\Delta E/E=22\%/\sqrt{E/10}.$~\citep{Hi1988}  Thus, the observed peak is consistent with being a 8 MeV $\bar{\nu}$ spectral line broadened by the estimated energy resolution.  

\subsection{Concerns about the $^8B$ data}
Six areas of potential concern about using the $^8B$ data to find the SN background are discussed, one of which leads to an adjustment in the background (dashed curve in Fig. 4(b)).  None of the six concerns is found to seriously undermine the validity of using the $^8B$ data, and in fact the one that leads to a background modiffication actually strengthens the case for the 7.5 MeV peak statistically.

\subsubsection{Time dependent background count rate}
The most obvious alternate explanation to a 7.5 MeV line in the data would be if the background count rate were time dependent, in which case the normalization of the SN and $^8B$ backgrounds need have no relation to one another.  A change was in fact made from a periodic cleaning of the tank water to a continuous cleaning following the SN, and during most of the $^8B$ data taking period.  However, ref.~\citep{Hi1988} notes that there was no time dependence in the background count rate for months prior to SN 1987A apart from small perturbations introduced by efforts to reduce the amount of Rn dissolved in the tank water.  Specifically, it was found that deviations from a constant rate were entirely consistent with a (random) Poisson distribution.  Thus, the small changes in radon level and background count rate due to the change from periodic to continuous cleaning would almost certainly have been small compared to the roughly $\sim 100\%$ increase in background that would be needed to explain the $\sim 600$ count excess in Fig. 4(b).   

\subsubsection{Fraction of events near the tank walls}
The larger fiducial mass for the SN data compared to the $^8B$ data has been adjusted for in finding the SN background.  However, that larger fiducial mass results in a much greater fraction of events near the detector walls that are caused by external radioactivity, and these events may have a different spectrum than centrally located events in the smaller $^8B$ fiducial mass.  It is noted in ref.~\citep{Hi1988}, however, that higher energy events that have $N_{hit}\ge 23$ ($E_\nu\ge 9.6$ MeV) are ``consistent with higher energy products of radioactivity at or outside the tank wall."  Thus, since events near the walls tend to be associated with energies above 9.6 MeV, while they might be responsible for the very small peak seen at 10 MeV in Fig. 5 (b) they could not account for the much larger peak centered at 7.5 MeV.

\subsubsection{Assumed background shape for $E_e<6$ MeV}
The $^8B$ data only showed a background spectrum for events having  $E_e>6$ MeV, so it is speculative to claim that background events having $E_e<6$ MeV should follow the same Gaussian functional form.  Support for this assertion can be found using data in ref.~\citep{Be2013} for the $^{214}Bi$ $\beta-$spectrum, which is the dominant background source for $E_e<9$ MeV.~\citep{Hi1988, Hi1989}  It may seem odd that a decay with a spectrum endpoint of only 3.27 MeV can be the dominant background for $E_e$ as high as 9 MeV, but in K-II one can easily have 2 or 3 PMT triggers in a given time window that mimic a single $e^\pm$ with $E_e$ as high as 10 MeV.  The upper grey curve in Fig. 6(a) shows a portion of the beta spectrum taken from Fig. 1 in ref.~\citep{Be2013} where the horizontal axis is the number of photoelectrons $P$ seen in 100 PMT's in a specific time interval.  $P$ is a measure of the energy, although it is not clear what proportionality constant to use to convert $P$ to $E_e$ or to $N_{hit}.$  Nevertheless, using the known efficiency function for the K-II detector (see ref.~\citep{Hi1988}) together with the measured spectrum one can deduce what K-II would measure for $^{214}Bi$ decays.  Using the choice of a proportionality constant $E_e=P/100$ (so as to yield about the right peak energy $E_e\approx 6.2$ MeV) one finds the predicted K-II background given by the dashed curve for the $^8B$ data.  The solid curve applies to the SN data, which as will be discussed has a different efficiency function.  This result can be seen to be consistent with a Gaussian on both sides of the maximum at around 6.2 MeV (p=620).  Thus, this exercise supports the use of a Gaussian (plus a small constant) to both $E_e<6$ MeV and $E_e>6$ MeV.   
\begin{figure}
\centerline{\includegraphics[angle=0,width=1.0\linewidth]{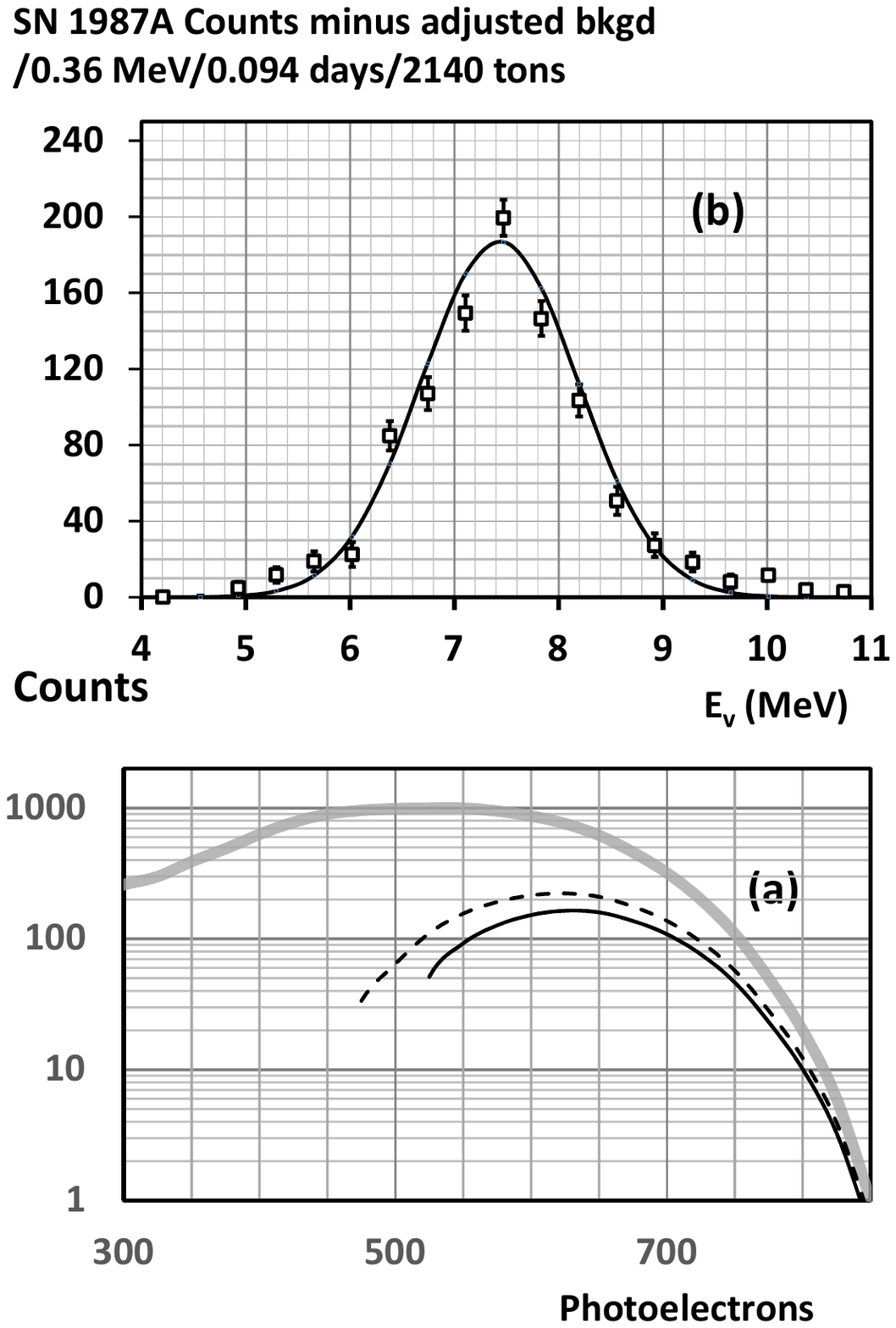}}
\caption{\label{fig4}(a) The grey curve is a portion of the measured spectrum of electrons in $\beta-$ decay of $^{214}Bi$ as depicted in ref.~\citep{Be2013}, while the black solid (SN data) and dashed ($^8B$ data) Gaussian-shaped curves were generated from the product of the measured spectrum and the K-II efficiency functions appropriate to the SN and $^8B$ data.   (b) Shows the background subtracted SN 1987A data using the adjusted background, together with a Gaussian fit.}
\end{figure}

\subsubsection{Different triggers in SN and $^8B$ data\label{triggers}}
Slightly different triggers were used for the SN and $^8B$ data sets.  
Thus, for the SN data an event required that $N_{hit}\ge 20$ within 100 nsec, but for the $^8B$ data this crterion was changed about halfway through the data-taking period to $N_{hit}\ge 18.$ 
This change affected the detector efficiency; thus for $N_{hit}\ge 20(18),$ 7.6- (6.7-) MeV electrons are detected with $50\%$ efficiency and 10- (8.8-) MeV electrons are detected with $90\%$ efficiency over the fiducial volume of the detector.  

The procedure that was used to adjust the K-II efficiency function so as to incorporate these differences is now described.  Fig. 3 of Ref.~\citep{Hi1988} shows a plot of the efficiency of the K-II detector versus energy for the case when the triggering criterion was $N_{hit}\ge 20$ in 100 ns.  Although the K-II authors do not provide a functional form for $\eta(E)$ versus E, it is found that for $E>E_0,$ the form: 
\begin{equation}
\eta(E)=1-e^{-\alpha(E-E_0)^{1.3}}
\end{equation}
describes this graph extremely well.  The two constants $E_0$ and $\alpha$ can easily be expressed in terms of the energies $E_{50}(E_{90})$ at which $\eta(E)=0.5(0.9),$ so that it is relatively easy to modify Eq. 5 so that it applies to the case $N_{hit}\ge 18$ in 100 ns instead of $N_{hit}\ge 20$ in 100 ns  for which those energies are 6.7(8.8) instead of 7.6(10) MeV.

Using the B-214 measured spectrum and the adjusted efficiency function, one finds that the expected background for the SN data compared to the $^8B$ data is: $26\%$ lower height, $10\%$ narrower width, and shifted to the right by $\Delta N_{hit}=+0.4.$ -- see Figs. 6(a) and 4(b).  If one uses the adjusted background for the SN data it is found that the background subtracted data is still consistent with a Gaussian centered on 7.5 MeV (see Fig. 5(b)).  Moreover, the excess above the adjusted background is obviously even more statistically significant than for the unadjusted background, and its broader width $(\Delta E/E=24\%$) is even more consistent with being a spectral line $(\Delta E/E=25\%$) than was found using the unadjusted background.
 
\subsubsection{Doubling of the trigger rate during the $^8B$ run}
It was found that when the change in triggering criterion was made about halfway though the data-taking period for the $^8B$ search the trigger rate doubled from 0.6 to 1.2 Hz.  If this doubled rate applied to half the data-taking period, it implies a background that is about $33\%$ lower for the SN data than the $^8B$ data.  In fact, based on the analysis in the previous section, the combination of reductions of $26\%$ in height and $10\%$ in width suggest a SN background rate that should be $33\%$ lower, which is in excellent agreement with the observed change in the count rate.  Thus, the average background count rate is unaffected after adjusting for the changed triggering criteria.

\subsubsection{Statistical significance of peak\label{error_bars}}
The null hypothesis is zero counts at all $E_\nu$ in Fig. 5(b).   According to Li and Ma~\citep{Li1983}, when the on-source (SN data) exposure time is negligible compared to the off-source ($^8B$ data) exposure time one may find the statistical significance for a single bin excess above background using:
\begin{equation}
S_i=\frac{N_{on}-\alpha N_{off}}{\sqrt{\alpha(N_{on}+N_{off})}}
\end{equation}
In the present case we have a very small on-source/off-source ratio ($\alpha=0.0007$), and since $N_B=\alpha N_{off},$ and $N_{on}<<N_{off},$ we have $S_i=(N_{on}-N_B)\sqrt{N_B}.$  Thus, the error bars for the $i^{th}$ bin in Fig. 4(b) and 5(b) are given by the square root of the background counts for that bin, and the $\chi^2$ for the null hypothesis is computed using $\chi^2=\sum S_i^2.$  One approach for dealing with bins having very small numbers of counts is to limit the sum to bins with say $N_{hit}\ge 10,$ for which we find $\chi^2=1022$ for 14 dof which yields a probability equivalent to a $30\sigma$ effect.   Had we used the adjusted background curve (see Fig. 4(b)) in doing the background subtraction, the result yields a slightly wider Gaussian (see Fig. 6(b)) with a statistical significance for the null hypothesis obviously $>30\sigma$.  Under the most conservative assumption where the data for $E_e<6$ MeV is completely ignored one obtains a $26\sigma$ effect ($\chi^2=695)$ for 5 dof.  In contrast to these null hypothesis cases, a fit of the background-subtracted data to a three parameter Gaussian spectral line is quite acceptable, and is shown in Figs. 5(b) and 6(b).  Thus, a best fit for Fig. 5(b) yields $\chi^2=20.1$ for the 13 central bins of the histogram, which for dof = 10 yields $p=8\%.$  Here since some bins have $N_{hit}< 10,$ we have used $\sqrt{N_B+1}$ in lieu of $\sqrt{N_B}.$\\

\subsection{Appendix conclusion}
The spectrum of $N\sim1000$ events recorded by the Kamiokande-II detector on the day of SN 1987A reveals an 8 MeV line (broadened by the expected $25\%$ resolution) atop a background found from other K-II data taken in the months after SN 1987A.  This $\bar{\nu}$ line, if genuine, has been well-hidden for 30 years because it occurs very close to the peak of the background. 
This fact might ordinarily justify extreme skepticism.  In the present case, however, a more positive view is called for based on 
(a) the very high statistical significance of the result $(30 \sigma),$ (b) the use of a detector background independent of the SN 1987A data, (c) the observed time-independence of the background count rate after adjusting for changed triggering criteria, and (d) the observation of an excess above the background spectrum whose central energy and width both agree with that of an 8 MeV $\bar{\nu}$ line broadened by $25\%$ resolution.  Most importantly, the last observation of an 8 MeV $\bar{\nu}$ line is in accord with the {\emph {prior prediction}} based on the Mont Blanc data, and the the dark matter model, itself supported by experimental observations.

\end{document}